\documentclass[11pt]{article}

\usepackage{acl}

\usepackage{times}
\usepackage{latexsym}

\usepackage[T1]{fontenc}
\usepackage[utf8]{inputenc}
\usepackage{microtype}
\usepackage{inconsolata}
\usepackage{graphicx}
\usepackage{amsmath}
\usepackage{amssymb}
\usepackage{booktabs}
\usepackage{caption}
\usepackage{subcaption}
\usepackage{float}

\usepackage[table,xcdraw]{xcolor}
\usepackage{makecell}
\usepackage{multirow}
\usepackage{mdframed}

\newcommand\blfootnote[1]{%
  \begingroup
  \renewcommand\thefootnote{}\footnote{#1}%
  \addtocounter{footnote}{-1}%
  \endgroup
}

\title{MARA: A Multimodal Adaptive Retrieval-Augmented Framework for Document Question Answering}

\author{
  Hui Wu\textsuperscript{1,2}\thanks{\ \ Both authors contributed equally to this research.}, 
  Haoquan Zhai\textsuperscript{2}\footnotemark[1], 
  Yuchen Li\textsuperscript{4},
  Hengyi Cai\textsuperscript{4},
  Peirong Zhang\textsuperscript{2}, \\
  \textbf{Yidan Zhang\textsuperscript{2},
  Lei Wang\textsuperscript{3}, 
  Chunle Wang\textsuperscript{3}, 
  Yingyan Hou\textsuperscript{3}\thanks{\ \ The corresponding author.}
  Shuaiqiang Wang\textsuperscript{4}, Dawei Yin\textsuperscript{4}}\\
  \textsuperscript{1}Key Laboratory of Target Cognition and Application Technology (TCAT), AIRI, CAS \\
  \textsuperscript{2}School of Electronic, Electrical and Communication Engineering, UCAS \\
  \textsuperscript{3}Aerospace Information Research Institute, Chinese Academy of Sciences \\
  \textsuperscript{4}Baidu Inc. \\
  \texttt{wuhui21@mails.ucas.ac.cn}, \texttt{zhqamazing@163.com}, \texttt{houyy@aircas.ac.cn}
}

\begin{document}
\maketitle
\blfootnote{Published in Proceedings of the 33rd ACM International Conference on Multimedia (MM '25), October 27--31, 2025, Dublin, Ireland.}

\begin{abstract}
 Retrieval-based multimodal document QA aims to identify and integrate relevant information from visually rich documents with complex multimodal structures. While retrieval-augmented generation (RAG) has shown strong performance in text-based QA, its extensions to multimodal documents remain underexplored and face significant limitations. Specifically, current approaches rely on query-agnostic document representations that overlook salient content and use static top-k evidence selection, which fails to adapt to the uncertain distribution of relevant information.
  To address these limitations, we propose the Multimodal Adaptive Retrieval-Augmented (MARA) framework, which introduces query-adaptive mechanisms to both retrieval and generation.
  MARA consists of two components: a Query-Aligned Region Encoder that builds multi-level document representations and reweights them based on query relevance to improve retrieval precision; and a Self-Reflective Evidence Controller that monitors evidence sufficiency during generation and adaptively incorporates content from lower-ranked sources using a sliding-window strategy. Experiments on six multimodal QA benchmarks demonstrate that MARA consistently improves retrieval relevance and answer quality over existing SOTA method.
\end{abstract}

\section{Introduction}
In real-world applications, documents increasingly exhibit multimodal characteristics by integrating text with structured layouts, tables, and visual elements \cite{xu2020layoutlm, li2021structurallm}. Such formats are commonly found in technical reports \cite{mathew2022infographicvqa}, academic papers \cite{li2024multimodal}, and presentation slides \cite{tanaka2023slidevqa}. Compared to plain-text documents, these formats contain richer and more complementary information across modalities \cite{zhai2023sigmoid}, offering stronger contextual grounding for downstream tasks such as question answering. As a result, retrieval-based multimodal document QA has attracted increasing attention, aiming to retrieve relevant information from diverse modalities to improve answer accuracy. A widely adopted approach for enhancing question answering with external information is retrieval-augmented generation (RAG) \cite{lewis2020retrieval}. RAG has demonstrated strong performance in open-domain and knowledge-intensive QA tasks \cite{yu2022retrieval,chen2024analyze,chen-etal-2025-terms}, particularly in text-only scenarios \cite{yasunaga2023retrieval, huang2024survey}. However, its application to multimodal documents remains underexplored. To bridge this gap, recent work has begun to extend RAG to incorporate visual information, such as VisRAG \cite{yu2024visrag}, which integrates visual features into both the retrieval and generation.

\begin{figure}[t]
    \centering
    \includegraphics[width=1\linewidth]{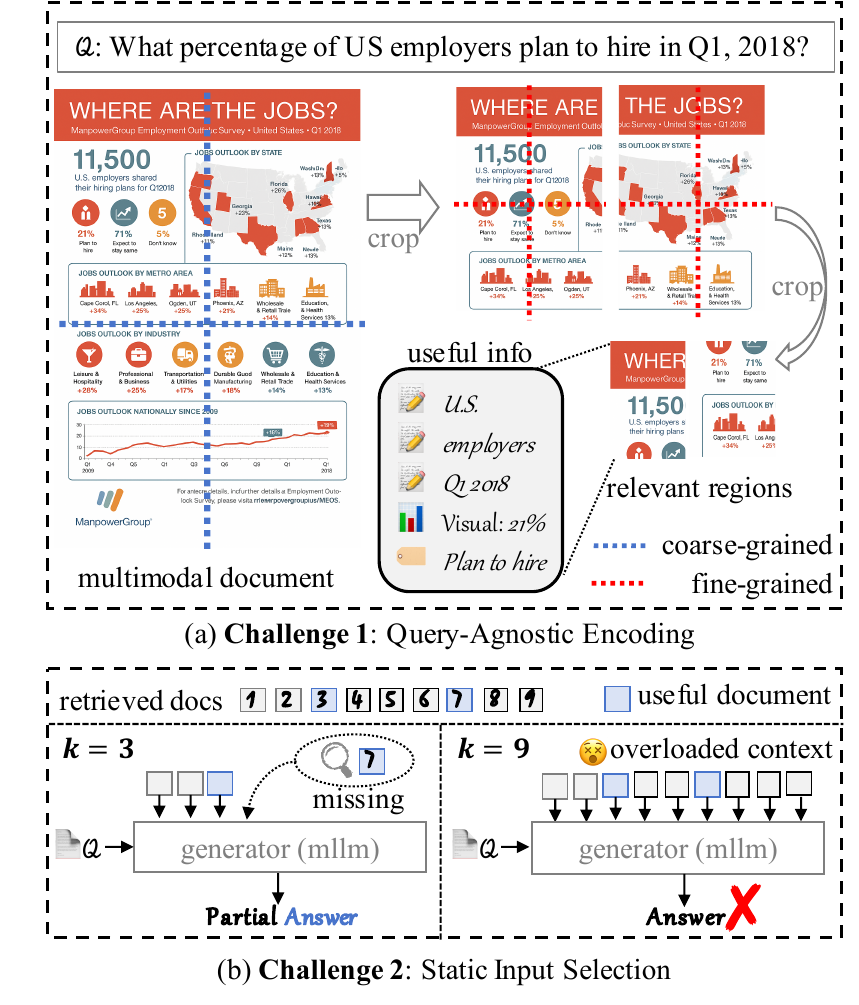}
    \caption{Two core challenges in multimodal document RAG. (a) Query-Agnostic Encoding: Without query-aware attention, both coarse and fine-grained regions may be overlooked. (b) Static Input Selection: The ranking position of useful document is unknown at inference time; fixed top-$k$ inputs may miss critical content or overload the generator with irrelevant information.}
    \label{fig1}
\end{figure}

However, current retrieval-augmented pipelines for multimodal document QA are inherently static, introducing two key limitations. 
First, the document encoder typically produces query-agnostic representations that overlook content most relevant to the question. 
This drawback becomes more pronounced in multimodal documents, where informative signals span from coarse layout structures to fine-grained visual cues.
As shown in Figure~\ref{fig1} (a), treating all regions equally leads to diluted attention: coarse-grained features lose specificity, while fine-grained lose context. 
Without query‑aware encoding, the retriever struggles to balance information across regions of different scales, resulting in inaccurate retrieval.

Besides, the retrieved documents are fed to the generator in a static, score-based order which leaves no room to adjust the number of input relevant documents.
Especially, the useful evidence may be clustered in a single document or dispersed across multiple sources with unknown distribution at inference time. 
As illustrated in Figure~\ref{fig1} (bottom), a small $k$ may miss essential content, yielding only a partial answer; a large $k$ introduces noise, overloading the generator’s context and degrading answer quality.This rigid selection thus fails to accommodate dynamic evidence patterns, often resulting in incomplete or ungrounded responses.

To address these limitations, we propose the Multimodal Adaptive Retrieval-Augmented (\textbf{MARA}) framework, which enhances both retrieval and generation with query-adaptive behaviors. Motivated by how humans selectively attend to relevant information and seek additional evidence when needed, MARA enables the system to focus on salient content and dynamically adjust the scope of information used based on the query. Specifically, the MARA framework consists of two components that introduce adaptivity into the retrieval and generation stages. On the retrieval side, the \textbf{Query-Aligned Region Encoder} constructs hierarchical document representations at global, coarse, and fine levels, and dynamically reweights subregions based on their alignment with the query. This design allows the retriever to focus on informative regions while preserving multi-scale document context. By aligning attention weights with query semantics, the model provides both improved retrieval accuracy and stronger interpretability. To support generation, the \textbf{Self-Reflective Evidence Controller} monitors evidence sufficiency during decoding and incrementally incorporates content from lower-ranked documents through a sliding-window mechanism. Rather than relying on a fixed top-k input, this adaptive evidence integration strategy balances coverage and noise, allowing the generator to retrieve complementary information when needed without overloading its context window. In summary, our contributions are as follows:
\vspace{-4pt}
\begin{itemize} 
    \item We introduce the Multimodal Adaptive Retrieval-Augmented (MARA) framework, which addresses the core limitations of query-agnostic retrieval and static evidence integration by incorporating query-adaptive mechanisms across both retrieval and generation stages.
    \item We propose Query-Aligned Region Encoding, a multi-granular representation module that dynamically reweights document regions based on query relevance. This enables fine-grained and interpretable retrieval in complex multimodal layouts.
    \item We develop the Self-Reflective Evidence Controller, which adaptively expands the evidence scope during generation by incrementally incorporating content from lower-ranked documents. This replaces fixed top-\textit{k} input strategies with a more flexible and context-aware approach.
    \item We conduct comprehensive experiments on six benchmark datasets spanning diverse domains and formats, demonstrating that MARA consistently improves both retrieval precision and answer quality compared to SOTA method.
\end{itemize}
\section{Related Work}

\subsection{Retrieval-based Multimodal Document QA}
 Retrieval-based multimodal document QA requires identifying and synthesizing relevant content from visually rich documents \cite{yu2024visrag,li2025m2oerank}.
 Unlike earlier approaches that assume the input document is provided \cite{methani2020plotqa,tanaka2023slidevqa}, this setting requires retrieving relevant content from a corpus prior to answering \cite{li2025towards, li2025rankexpert}. Benchmark datasets such as DocVQA \cite{tito2023hierarchical}, InfographicVQA \cite{mathew2022infographicvqa}, and ChartQA \cite{xu2023chartbench} were initially introduced for non-retrieval QA but have recently been adapted for retrieval-based scenarios to better reflect real-world applications \cite{yu2024visrag}.
 To support such settings, retrieval-augmented generation (RAG) has become a standard framework for improves generation quality by grounding answers in external content \cite{lewis2020retrieval, yasunaga2023retrieval, huang2024survey, ram2023context, shi2024replug}. Early RAG systems focused on retrieving text passages from corpora such as Wikipedia \cite{wang2023self, ma2023query, li2025rankelectra}, and have since been extended to incorporate citation grounding \cite{gao2023enabling,wang2023survey} and structured knowledge sources \cite{edge2024local,xiong2024search,sarmah2024hybridrag}. More recent efforts explore multimodal retrieval, such as retrieving text–image pairs for open-domain QA \cite{chang2022webqa, luo2023end}, highlighting growing interest in cross-modal grounding.
 However, the application of RAG to multimodal document QA remains underexplored. Compared to open-domain QA or image-based tasks, this setting introduces unique demands across both retrieval and generation stages, calling for adaptive strategies that can better align information selection and integration with query-specific requirements.

\subsection{Document Representation for Retrieval}
Early methods for multimodal document representation focus on joint modeling of text, layout, and visual information using Transformers such as LayoutLMv2 \cite{xu2021layoutlmv2}, DocFormer \cite{appalaraju2021docformer}, and Donut \cite{kim2022ocr}. To support retrieval-based QA, recent work leverages vision-language models to encode entire documents and designs contrastive objectives that align document and query embeddings in a shared space, as in SigLIP \cite{zhai2023sigmoid} and DSE \cite{ma2024unifying}. While effective at global retrieval, these approaches produce static, document-level embeddings that lack fine-grained control. In parallel, ColPali \cite{faysse2024colpali} computes token-level similarity scores during retrieval, but does not modify the underlying document representation. In contrast, we focus on the encoding stage and propose a multi-granular, query-sensitive strategy that dynamically reweights document regions based on their relevance, enabling adaptive and fine-grained retrieval within RAG.

\subsection{Multimodal Generation}
\begin{figure*}[t!]
    \centering
    \includegraphics[width=\linewidth]{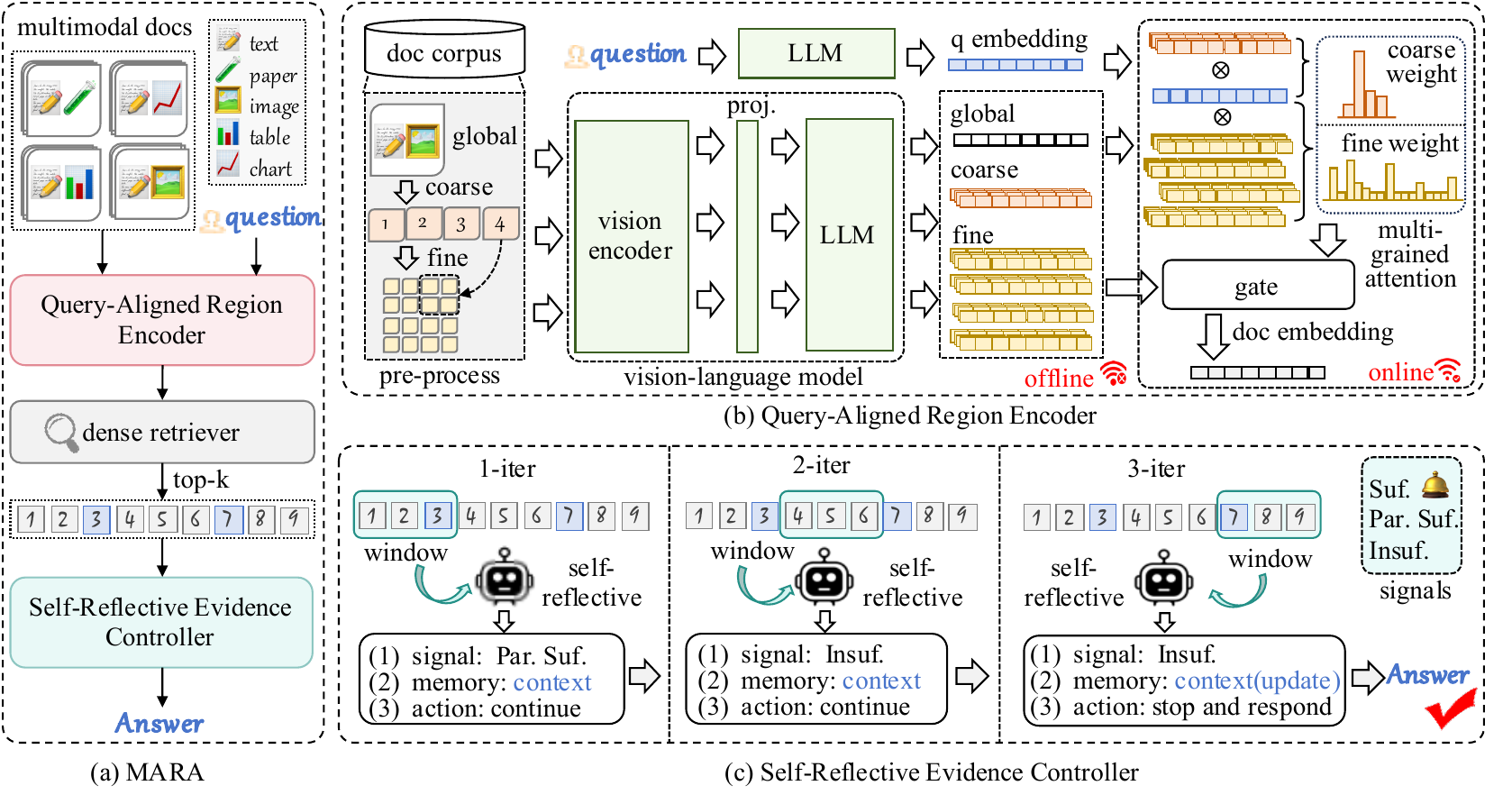}
    \caption{Overview of the proposed MARA framework. (a) MARA adaptively retrieves and integrates evidence from multimodal documents for question answering. (b) The Query-Aligned Region Encoder constructs multi-granular document embeddings by reweighting global, coarse, and fine-level features according to query relevance. (c) The Self-Reflective Evidence Controller incrementally incorporates documents through a sliding window, using a self-monitoring mechanism to assess evidence sufficiency during generation.}
    \label{fig2}
\end{figure*}
Multimodal RAG systems increasingly employ large vision-language models as generators to decode answers from retrieved documents \cite{zhang2024mm,li2023s2phere}. The generator typically operates over a fixed top-k set of inputs and lack the ability to assess whether the provided evidence is sufficient. To address this limitation, recent work explores static input selection strategies such as prompt optimization \cite{cuconasu2024power}, document reranking \cite{yu2024rankrag}, and compressive context selection \cite{xu2024recomp}. For example, COMPACT \cite{yoon2024compact} incrementally compresses retrieved segments and relies on external evaluation to determine sufficiency. However, these methods remain detached from the generation process and offer limited adaptability. In contrast, we investigate generation-time feedback as a mechanism to guide evidence usage in a more flexible and context-aware manner.

\section{Method}

\subsection{Task Formulation}

Given a natural language query $q$ and a corpus of multimodal documents $\mathcal{D} = \{x_j\}_{j=1}^N$, where each document $x_j$ contains text, layout, tables, and visual elements, the goal is to generate a textual answer $a$ grounded in relevant content.

Following the retrieval-augmented generation (RAG) framework, the task proceeds in two stages:
\begin{itemize}
    \item \textbf{Retrieval.} A multimodal encoder $f_\theta$ ranks documents in $\mathcal{D}$ by their semantic relevance to $q$, retrieving a top-$k$ subset $\mathcal{R}_k = \{x^{(1)}, \dots, x^{(k)}\}$.
    \item \textbf{Generation.} A decoder $G_\theta$ generates an answer $a = G_\theta(q, \mathcal{R}_k)$ based on the query and retrieved content.
\end{itemize}

This formulation introduces two key challenges emphasized in Figure~\ref{fig1}: the need for query-aligned encoding across hierarchical regions, and the dynamic nature of evidence distribution. Our MARA framework addresses both with adaptive mechanisms in retrieval and generation. An overview of the framework is shown in Figure~\ref{fig2}, and we describe its two main components in the following sections.

\subsection{Query-Aligned Region Encoder}
In multimodal document QA, information relevant to a query may appear at vastly different structural scales, from high-level layouts such as section headers or figure blocks to fine-grained elements like table cells or localized image regions. Traditional document encoders tend to produce fixed, query-agnostic representations, often overlooking subtle but critical content. To address this, we propose the \textbf{Query-Aligned Region Encoder}, which constructs multi-level representations of each document in an offline stage, and dynamically fuses them with query-aligned signals during inference, as illustrated in Figure~\ref{fig2}(b).

Given a document $x_j$, we first construct a hierarchical structural decomposition at three granularities. The \textit{global level} considers the entire document as a whole; the \textit{coarse level} partitions it into $m$ high-level regions $\{x_j^{(i)}\}_{i=1}^m$ such as layout blocks or composite visual-textual segments; and the \textit{fine level} further divides each coarse region into $k$ smaller patches $\{x_j^{(i,l)}\}_{l=1}^k$, capturing local structures like phrase spans, table cells, or image crops.

All regions are independently encoded by a shared multimodal encoder $f_\theta$. This yields three sets of embeddings:
\begin{align}
    \mathcal{E}_j^g &= \left\{f_\theta(x_j)\right\}, \\
    \mathcal{E}_j^c &= \left\{f_\theta(x_j^{(i)})\right\}_{i=1}^{m}, \\
    \mathcal{E}_j^f &= \left\{f_\theta(x_j^{(i,l)})\right\}_{i=1, l=1}^{m, k}.
\end{align}

Here, $\mathcal{E}_j^g$ denotes the global-level embedding, while $\mathcal{E}_j^c$ and $\mathcal{E}_j^f$ represent the coarse and fine region embeddings, respectively. These three components are unified as the document's multi-granular representation:
\[
\mathcal{E}_j = \mathcal{E}_j^g \cup \mathcal{E}_j^c \cup \mathcal{E}_j^f.
\]

This entire decomposition and encoding process is performed independently of the query and cached offline, enabling efficient indexing and retrieval at inference time.

At inference time, we align the offline representations with the input query to produce a query-aligned document embedding. This process involves two steps: query-to-region attention and adaptive feature fusion.

We first compute the relevance between the query $q$ and each region embedding at the coarse and fine levels using scaled dot-product attention. Specifically, let $f_\theta(q)$ denote the query embedding. For each granularity $s \in \{c, f\}$, we compute attention weights over the corresponding embedding set $\mathcal{E}_j^s$ as:
\begin{equation}
\alpha^s = \mathrm{Softmax}\left( \frac{f_\theta(q)^\top \mathcal{E}_j^s}{\tau} \right),
\end{equation}
where $\tau$ is a learnable temperature parameter. The attended feature for each level is then given by:
\begin{equation}
\mathcal{E}_j^{s,\mathrm{att}} = \sum_{k=1}^{|\mathcal{E}_j^s|} \alpha_k^s \cdot \mathcal{E}_j^s(k).
\end{equation}

These attention maps highlight the most query-relevant subregions within the document’s structural hierarchy, allowing the model to focus on content at appropriate granularity.

Next, we adaptively combine the global, coarse, and fine representations into a single query-aligned embedding via gated fusion. A lightweight feedforward module predicts the fusion weights $[g_c, g_f]$ from the query and aggregated document features:
\begin{equation}
[g_c, g_f] = \sigma\left( W[f_\theta(q);\mathrm{Agg}(\mathcal{E}_j)] \right),
\end{equation}
where $\sigma$ is the sigmoid activation and $\mathrm{Agg}(\cdot)$ denotes feature pooling (e.g., mean) over all region embeddings. The final document embedding is computed as:
\begin{equation}
\mathcal{E}_j' = g_c \cdot \mathcal{E}_j^{c,\mathrm{att}} + g_f \cdot \mathcal{E}_j^{f,\mathrm{att}} + (1 - g_c - g_f) \cdot \mathcal{E}_j^g.
\end{equation}

This design allows all hierarchical representations to be precomputed offline, enabling efficient indexing and retrieval. At inference time, query-aligned attention and gated fusion guide the model to emphasize either layout-level or fine-grained content based on the query, while preserving the global context. By decoupling document encoding from query interaction, QRE achieves both computational efficiency and semantic adaptivity, which is critical for scalable multimodal document QA.

The final similarity score between $x_j$ and $q$ is computed via cosine similarity:
\begin{equation}
s_j = \frac{f_\theta(q)^\top \mathcal{E}_j'}{\|f_\theta(q)\| \cdot \|\mathcal{E}_j'\|},
\end{equation}
and the top-$k$ documents are selected for generation accordingly.

\subsection{Self-Reflective Evidence Controller}

A key limitation of traditional RAG pipelines lies in their static input strategy: a fixed top-$k$ set of documents is passed to the generator, regardless of query complexity or evidence distribution. As discussed earlier, such rigid selection often leads to incomplete answers when $k$ is too small, or noisy and unfocused outputs when $k$ is too large. To overcome this issue, we propose the \textbf{Self-Reflective Evidence Controller}, a generation-time module that enables the model to expand its input context through iterative feedback.

As illustrated in Figure~\ref{fig2}(c), the controller operates over the top-$k$ retrieved candidates $\mathcal{R}_k = \{x^{(1)}, \dots, x^{(k)}\}$ using a fixed-size sliding window $\mathcal{R}_t^{(i)}$ of size $t$. At each step $i$, the generator is prompted to reflect on the current evidence, which consists of the current window and an accumulated \textit{context memory} $\mathcal{M}^{(i-1)}$ containing retained information from previous steps. The generator receives this input along with an in-context prompt $C_\text{signal}$ and produces a textual judgment, which is parsed into a discrete sufficiency signal:

\begin{equation}
s^{(i)} = \mathrm{ParseSignal}\left(G_\theta(q, \mathcal{R}_t^{(i)} \cup \mathcal{M}^{(i-1)}, C_\text{signal})\right),
\end{equation}
where $s^{(i)} \in \{\texttt{Suf.}, \texttt{Par.suf.}, \texttt{Insuf.}\}$.

If the model returns \texttt{Suf.}, it proceeds to generate the final answer:

\begin{equation}
a^{(i)} \leftarrow G_\theta(q, \mathcal{R}_t^{(i)} \cup \mathcal{M}^{(i-1)}).
\end{equation}

If the context is deemed \texttt{Par.suf.}, the model extracts useful content from the current window and appends it to the context memory:

\begin{equation}
\mathcal{M}^{(i)} \leftarrow \mathcal{M}^{(i-1)} \cup \mathrm{ExtractRelevant}(\mathcal{R}_t^{(i)}),
\end{equation}

and continues to the next iteration.

If the signal is \texttt{Insuf.}, the window simply advances without memory update.

Optionally, the generator may also return a structured feedback trace to guide subsequent retrieval steps:

\begin{equation}
f^{(i)} \leftarrow G_\theta(q, \mathcal{R}_t^{(i)} \cup \mathcal{M}^{(i-1)}, C_\text{guide}),
\end{equation}

where $C_\text{guide}$ prompts the model to indicate what additional information is required.
\begin{table*}[!t]
    \centering
    \caption{Overall generation performance compared to VisRAG in accuracy (\%). Performance relative to the Oracle (using the ground-truth document(s) for generation) is shown in {\color{blue}blue}, with the best result highlighted in \textbf{bold}.}
    \label{table1}
    \resizebox{0.99\linewidth}{!}{
    \begin{tabular}{ll|l|l|l|l|l|l|l}
    \hline
    \textbf{Model} & \textbf{Input} & \textbf{ArxivQA} & \textbf{ChartQA} & \textbf{DocVQA} & \textbf{InfoVQA} & \textbf{PlotQA} & \textbf{SlideVQA} & \textbf{Average} \\
    \hline
    \rowcolor{gray!15}
    \multicolumn{9}{c}{\textbf{(a) VisRAG}} \\
    \hline
    \multirow{4}{*}{\makecell[l]{MiniCPM-V 2.6}} 
    & top-1 & 65.39 {\small\textcolor{blue}{(86.0\%)}} & 44.71 {\small\textcolor{blue}{(66.3\%)}} & 69.40 {\small\textcolor{blue}{(77.6\%)}} & 49.91 {\small\textcolor{blue}{(72.5\%)}} & 42.89 {\small\textcolor{blue}{(79.7\%)}} & 62.44 {\small\textcolor{blue}{(84.8\%)}} & 55.79 {\small\textcolor{blue}{(78.0\%)}} \\
    & top-3 & 65.76 {\small\textcolor{blue}{(86.5\%)}} & 42.90 {\small\textcolor{blue}{(63.6\%)}} & 74.50 {\small\textcolor{blue}{(83.3\%)}} & 50.94 {\small\textcolor{blue}{(74.0\%)}} & 37.57 {\small\textcolor{blue}{(69.8\%)}} & 67.87 {\small\textcolor{blue}{(92.1\%)}} & 56.59 {\small\textcolor{blue}{(79.1\%)}} \\
    & top-10 & 64.78 {\small\textcolor{blue}{(85.2\%)}} & 44.01 {\small\textcolor{blue}{(65.3\%)}} & 59.07 {\small\textcolor{blue}{(66.1\%)}} & 43.20 {\small\textcolor{blue}{(62.8\%)}} & 35.12 {\small\textcolor{blue}{(65.3\%)}} & 67.80 {\small\textcolor{blue}{(92.0\%)}} & 52.33 {\small\textcolor{blue}{(73.2\%)}} \\
    & Oracle & \textcolor{gray}{76.03 (100\%)} & \textcolor{gray}{67.41 (100\%)} & \textcolor{gray}{89.41 (100\%)} & \textcolor{gray}{68.80 (100\%)} & \textcolor{gray}{53.79 (100\%)} & \textcolor{gray}{73.66 (100\%)} & \textcolor{gray}{71.52 (100\%)} \\
    \hline
    \multirow{3}{*}{GPT-4o} 
    & top-1 & 62.94 {\small\textcolor{blue}{(93.1\%)}} & 43.66 {\small\textcolor{blue}{(64.0\%)}} & 68.32 {\small\textcolor{blue}{(77.5\%)}} & 55.32 {\small\textcolor{blue}{(76.2\%)}} & 33.73 {\small\textcolor{blue}{(76.2\%)}} & 64.74 {\small\textcolor{blue}{(75.8\%)}} & 54.79 {\small\textcolor{blue}{(77.1\%)}} \\
    & top-3 & 62.26 {\small\textcolor{blue}{(92.1\%)}} & 46.52 {\small\textcolor{blue}{(68.2\%)}} & 78.21 {\small\textcolor{blue}{(88.8\%)}} & 58.27 {\small\textcolor{blue}{(80.2\%)}} & 30.12 {\small\textcolor{blue}{(68.1\%)}} & 69.41 {\small\textcolor{blue}{(81.3\%)}} & 57.47 {\small\textcolor{blue}{(80.9\%)}} \\
    & Oracle & \textcolor{gray}{67.58 (100\%)} & \textcolor{gray}{68.24 (100\%)} & \textcolor{gray}{88.10 (100\%)} & \textcolor{gray}{72.64 (100\%)} & \textcolor{gray}{44.25 (100\%)} & \textcolor{gray}{85.42 (100\%)} & \textcolor{gray}{71.04 (100\%)} \\
    \hline
    \rowcolor{gray!15}
    \multicolumn{9}{c}{\textbf{(b) Ours}} \\
    \hline
    \multirow{5}{*}{MiniCPM-V 2.6} 
    & top-1 & 69.47 {\small\textcolor{blue}{(91.4\%)}} & 44.43 {\small\textcolor{blue}{(65.9\%)}} & 69.45 {\small\textcolor{blue}{(77.7\%)}} & 51.86 {\small\textcolor{blue}{(75.4\%)}} & 42.89 {\small\textcolor{blue}{(79.7\%)}} & 62.68 {\small\textcolor{blue}{(85.1\%)}} & 56.80 {\small\textcolor{blue}{(79.4\%)}} \\
    & top-3 & 70.07 {\small\textcolor{blue}{(92.2\%)}} & 44.57 {\small\textcolor{blue}{(66.1\%)}} & 76.53 {\small\textcolor{blue}{(85.6\%)}} & 52.20 {\small\textcolor{blue}{(75.9\%)}} & 40.05 {\small\textcolor{blue}{(74.5\%)}} & 66.40 {\small\textcolor{blue}{(90.1\%)}} & 58.30 {\small\textcolor{blue}{(81.5\%)}} \\
    & top-10 & 69.65 {\small\textcolor{blue}{(91.6\%)}} & 41.64 {\small\textcolor{blue}{(61.8\%)}} & 60.56 {\small\textcolor{blue}{(67.7\%)}} & 37.73 {\small\textcolor{blue}{(54.8\%)}} & 32.96 {\small\textcolor{blue}{(61.3\%)}} & 67.56 {\small\textcolor{blue}{(91.7\%)}} & 51.68 {\small\textcolor{blue}{(72.3\%)}} \\
    & \textbf{MARA} & \textbf{71.85} {\small\textcolor{blue}{(94.5\%)}} & \textbf{53.64} {\small\textcolor{blue}{(79.6\%)}} & \textbf{80.59} {\small\textcolor{blue}{(90.1\%)}} & \textbf{58.07} {\small\textcolor{blue}{(84.4\%)}} & \textbf{44.57} {\small\textcolor{blue}{(82.9\%)}} & \textbf{69.48} {\small\textcolor{blue}{(94.3\%)}} & \textbf{63.03} {\small\textcolor{blue}{(88.1\%)}} \\
    & Oracle & \textcolor{gray}{76.03 (100\%)} & \textcolor{gray}{67.41 (100\%)} & \textcolor{gray}{89.41 (100\%)} & \textcolor{gray}{68.80 (100\%)} & \textcolor{gray}{53.79 (100\%)} & \textcolor{gray}{73.66 (100\%)} & \textcolor{gray}{71.52 (100\%)} \\
    \hline
    \multirow{4}{*}{GPT-4o} 
    & top-1 & 63.19 {\small\textcolor{blue}{(93.5\%)}} & 44.17 {\small\textcolor{blue}{(64.7\%)}} & 70.28 {\small\textcolor{blue}{(79.8\%)}} & 58.93 {\small\textcolor{blue}{(81.1\%)}} & 33.41 {\small\textcolor{blue}{(72.5\%)}} & 67.79 {\small\textcolor{blue}{(79.4\%)}} & 56.30 {\small\textcolor{blue}{(79.2\%)}} \\
    & top-3 & 63.90 {\small\textcolor{blue}{(94.6\%)}} & 45.89 {\small\textcolor{blue}{(67.2\%)}} & 81.22 {\small\textcolor{blue}{(92.2\%)}} & 62.35 {\small\textcolor{blue}{(85.8\%)}} & 32.18 {\small\textcolor{blue}{(72.7\%)}} & 69.20 {\small\textcolor{blue}{(81.0\%)}} & 59.12 {\small\textcolor{blue}{(83.2\%)}} \\
    & \textbf{MARA} & \textbf{63.92} {\small\textcolor{blue}{(94.6\%)}} & \textbf{51.25} {\small\textcolor{blue}{(75.1\%)}} & \textbf{84.64} {\small\textcolor{blue}{(96.1\%)}} & \textbf{68.02} {\small\textcolor{blue}{(93.6\%)}} & \textbf{32.98} {\small\textcolor{blue}{(74.5\%)}} & \textbf{73.40} {\small\textcolor{blue}{(85.9\%)}} & \textbf{62.37} {\small\textcolor{blue}{(87.8\%)}} \\
    & Oracle & \textcolor{gray}{67.58 (100\%)} & \textcolor{gray}{68.24 (100\%)} & \textcolor{gray}{88.10 (100\%)} & \textcolor{gray}{72.64 (100\%)} & \textcolor{gray}{44.25 (100\%)} & \textcolor{gray}{85.42 (100\%)} & \textcolor{gray}{71.04 (100\%)} \\
    \hline
    \end{tabular}
    }
    \label{table1}
\end{table*}

The process terminates as soon as a \texttt{Suf.} signal is observed, enabling early stopping. If all candidates in $\mathcal{R}_k$ are exhausted without reaching sufficiency, the generator falls back to memory to produce a final output:

\begin{equation}
a_{\text{final}} =
\begin{cases}
G_\theta(q, \mathcal{M}^{(i)}), & \text{if } |\mathcal{M}^{(i)}| > 0, \\
\texttt{abstain}, & \text{otherwise}.
\end{cases}
\end{equation}

This self-reflective mechanism allows the generator to actively manage its input scope, reducing unnecessary context while preserving critical signals. Unlike prior methods that rely on fixed thresholds or auxiliary classifiers, SEC enables more robust and context-aware generation under varied evidence distributions.

Together, the retrieval encoder and the generation controller form a dual-adaptive framework for multimodal document question answering. The retrieval module selects content with hierarchical, query-aware attention, while the generation module incrementally adjusts its input based on evidence sufficiency. This coordinated design enables the system to align information scope with reasoning needs in a dynamic and efficient manner.
\section{Experiments}
\subsection{Experimental Setup}

\subsubsection{Datasets}
We conduct evaluations on six publicly available benchmarks that span a broad spectrum of multimodal document question answering scenarios. These datasets vary in source domain, visual layout, and modality composition:

\begin{itemize}
    \item \textbf{MP-DocVQA}~\cite{tito2023hierarchical}: A dataset consisting of scanned industrial manuals paired with natural language questions.

    \item \textbf{ArXivQA}~\cite{li2024multimodal}: Comprises scientific publications from arXiv, including full-text PDFs with structured elements such as sections, equations, tables, and embedded figures.

    \item \textbf{ChartQA}~\cite{xu2023chartbench}: Focuses on bar chart-based questions. Each instance includes an image of a chart and associated metadata in table format, designed for visual-textual alignment and fact retrieval.

    \item \textbf{PlotQA}~\cite{methani2020plotqa}: Similar in format to ChartQA but centered on line charts. The dataset includes synthetic and real-world plots with numerical queries based on trends, comparisons, or specific data points.

    \item \textbf{InfographicsVQA}~\cite{mathew2022infographicvqa}: Contains infographic images that combine textual and symbolic information.

    \item \textbf{SlideVQA}~\cite{tanaka2023slidevqa}: A collection of slide decks accompanied by multi-turn QA pairs. 
\end{itemize}

These benchmarks collectively cover a diverse set of document types, including scientific articles, scanned pages, data visualizations, and presentation slides. We follow the official data splits for each benchmark and report evaluation results on the corresponding test sets. Dataset statistics and preprocessing details are provided in the Appendix.

\subsubsection{Baselines}
We compare our method with representative baselines from two categories. 

\noindent
\textbf{Document-level generation.} We use \textbf{VisRAG}~\cite{yu2024visrag}\footnote{\url{https://github.com/OpenBMB/VisRAG}} as our primary baseline. It represents a recent multimodal RAG pipeline that integrates visual-language modeling with RAG.

\begin{table*}[!t]
\centering
\caption{Retrieval performance compared to VLM-Models in MRR and Recall. The best result is highlighted in \textbf{bold}.}
\label{retrieve_baseline}
\resizebox{0.99\linewidth}{!}{
\begin{tabular}{l|c|c|c|c|c|c|c|c|c|c|c|c|c|c}
\hline
\multirow{2}{*}{\textbf{Method}} & 
\multicolumn{2}{c}{\textbf{ArxivQA}} & 
\multicolumn{2}{c}{\textbf{ChartQA}} & 
\multicolumn{2}{c}{\textbf{DocVQA}} &
\multicolumn{2}{c}{\textbf{InfoVQA}} & 
\multicolumn{2}{c}{\textbf{PlotQA}} & 
\multicolumn{2}{c}{\textbf{SlideVQA}} &
\multicolumn{2}{c}{\textbf{Average}} \\
\cmidrule(lr){2-3} \cmidrule(lr){4-5} \cmidrule(lr){6-7} \cmidrule(lr){8-9} 
\cmidrule(lr){10-11} \cmidrule(lr){12-13} \cmidrule(lr){14-15}
& MRR & Recall & MRR & Recall & MRR & Recall & MRR & Recall & MRR & Recall & MRR & Recall & MRR & Recall \\
\hline
\rowcolor{gray!15}
\multicolumn{15}{c}{\textbf{(a) VLM-models}} \\
\hline
SigLIP & 50.22 & 49.51 & 61.44 & 63.65 & 66.01 & 78.61 & 72.87 & 82.16 & 40.15 & 41.36 & 89.51 & 93.03 & 63.37 & 68.05 \\
DSE & 63.76 & 77.57 & \textbf{62.68} & 74.51 & 81.55 & 93.88 & 85.39 & 96.58 & \textbf{45.70} & \textbf{64.22} & 93.84 & 97.20 & 72.16 & 83.99 \\
ColPali & 64.36 & 76.46 & 56.46 & 70.19 & \textbf{84.10} & 96.06 & 78.74 & 93.65 & 30.37 & 48.62 & \textbf{93.91} & 96.97 & 67.99 & 80.33 \\
VisRAG & 66.70 & 80.42 & 58.74 & 72.42 & 77.64 & 92.92 & 84.12 & 96.24 & 40.19 & 61.40 & 91.76 & 96.86 & 69.86 & 83.38 \\
\hline
\rowcolor{gray!15}
\multicolumn{15}{c}{\textbf{(b) Ours}} \\
\hline
QRE-HF & 69.00 & 81.99 & 59.71 & 73.96 & 81.69 & 95.16 & 86.99 & 97.61 & 39.90 & 61.16 & 93.57 & 97.60 & 71.81 & 83.91 \\
QRE-Rep. & \textbf{72.02} & \textbf{85.22} & 60.36 & \textbf{75.63} & 83.35 & \textbf{96.59} & \textbf{87.93} & \textbf{98.04} & 43.23 & 63.87 & 93.90 & \textbf{97.73} & \textbf{73.46} & \textbf{86.18} \\
\hline
\end{tabular}
}
\label{table2}
\end{table*}
\noindent
\textbf{Retrieval.} We consider several multi-vector document encoders trained to produce dense embeddings from page images. This includes:
\begin{itemize}
    \item \textbf{SigLIP}~\cite{zhai2023sigmoid}\footnote{\url{https://huggingface.co/HuggingFaceM4/siglip-so400m-14-980-flash-attn2-navit}}: a CLIP-style vision model;
    \item \textbf{ColPali}~\cite{faysse2024colpali}\footnote{\url{https://github.com/illuin-tech/colpali?tab=readme-ov-file}}: a multi-vector retriever for document layout;
    \item \textbf{DSE}~\cite{ma2024fine}\footnote{\url{https://github.com/texttron/tevatron}} and \textbf{VisRAG-Ret}~\cite{yu2024visrag}: representing the current state-of-the-art in visual document embedding.
\end{itemize}

\subsubsection{Evaluation Metrics}
To evaluate generation quality, we use \textbf{MiniCPM3-4B} \cite{hu2024minicpm} as an automatic evaluator, which captures semantic equivalence beyond surface forms and better aligns with human judgment than metrics like BLEU or exact match. This choice is consistent with prior work showing that semantically-aware evaluation is more reliable for short factual answers with formatting variation. For retrieval, we report \textbf{MRR@10} and \textbf{Recall@10}, which measure whether relevant documents appear among the top retrieved candidates.

\subsubsection{Implementation Details}
Our Ad-VisRAG framework consists of a retrieval encoder and a generative model. For retrieval, we evaluate both the publicly released VisRAG-Ret checkpoint\footnote{\url{https://huggingface.co/openbmb/VisRAG-Ret}} and a domain-adapted variant finetuned on each dataset using 8$\times$A100 GPUs with a batch size of 128. For generation, we use \textbf{MiniCPM-V 2.6} \cite{ma2024fine} and \textbf{GPT-4o} accessed via API. Complete hyperparameter configurations are provided in Appendix.
\subsection{Main Results}
In this section, we present experiments on both overall generation and retrieval performance.

\noindent\textbf{Generation Result.} 
Table~\ref{table1} reports generation accuracy across six datasets, comparing MARA framework with VisRAG and Oracle settings. MARA consistently outperforms VisRAG under both MiniCPM-V 2.6 and GPT-4o backbones, achieving 88.1\% and 87.8\% of Oracle performance on average, respectively. These results demonstrate that the two core modules jointly enable MARA to approach Oracle-level performance without requiring exhaustive input.

Notably, increasing the number of retrieved documents (e.g., from top-3 to top-10) does not necessarily improve performance. On datasets such as PlotQA, additional documents often introduce irrelevant content, which distracts the generator and reduces answer accuracy. This highlights the limitation of static top-$k$ selection strategies that cannot assess evidence sufficiency or filter contextual noise in real time. In contrast, our \textit{Self-Reflective Evidence Controller} (SEC) dynamically expands evidence through a feedback-guided sliding window and memory mechanism, allowing the model to selectively integrate useful information. This yields more concise and accurate answers and is particularly advantageous for tasks with heterogeneous evidence distributions.

Performance differences across tasks also reflect the varying modality and reasoning demands. For instance, PlotQA requires precise numerical reasoning over visual plots, which poses challenges even under Oracle input. While the absolute scores remain lower, Ad-VisRAG consistently narrows the gap to Oracle, indicating that generation quality depends not only on retrieval coverage but also on the model’s ability to regulate its input based on context sufficiency. This further highlights the contribution of the SEC module in stabilizing generation across task types.

\noindent\textbf{Retrieval Results.}
We evaluate our proposed \textit{Query-Aligned Region Encoder} (QRE) in two variants: a publicly released checkpoint (QRE-HF) and an in-domain finetuned version (QRE-Rep). As shown in Table~\ref{table2}, both variants consistently outperform state-of-the-art VLM-based retrievers on all datasets. Compared to VisRAG-Ret, QRE-Rep yields average gains of 3.6 MRR and 2.8 Recall points, with the largest improvements on PlotQA and SlideVQA.

These results demonstrate the effectiveness of our multi-granular, query-aware encoding in capturing relevant visual-textual cues. Consistent performance across document types, from scientific papers to complex plots, underscores the robustness and generalizability of QRE as a drop-in retriever for multimodal QA. We further analyze the impact of document granularity and attention temperature ($\tau$) on this improvement in the following section.

\subsection{Ablation Study}
\begin{table}[htbp]
\centering
\caption{Ablation study of MARA. Performance relative to the Oracle is shown in {\color{blue}blue}; best results are in \textbf{bold}.}
\label{table3}
\resizebox{0.95\linewidth}{!}{
\begin{tabular}{l|l|l|l}
\hline
\textbf{Model} & \textbf{DocVQA} & \textbf{InfoVQA} & \textbf{PlotQA} \\
\hline
\rowcolor{gray!15}
\multicolumn{4}{c}{\textbf{MiniCPM-V 2.6}} \\
\hline
\textbf{MARA} & \textbf{80.59} {\small\textcolor{blue}{(90.1\%)}} & \textbf{58.07} {\small\textcolor{blue}{(84.4\%)}} & \textbf{44.57} {\small\textcolor{blue}{(82.9\%)}} \\
w/o SEC & 76.53 {\small\textcolor{blue}{(85.6\%)}} & 52.20 {\small\textcolor{blue}{(75.9\%)}} & 40.05 {\small\textcolor{blue}{(74.5\%)}} \\
w/o QRE & 74.95 {\small\textcolor{blue}{(83.3\%)}} & 51.76 {\small\textcolor{blue}{(75.2\%)}} & 41.16 {\small\textcolor{blue}{(76.5\%)}} \\
w/o both & 74.50 {\small\textcolor{blue}{(83.3\%)}} & 50.94 {\small\textcolor{blue}{(70.0\%)}} & 37.57 {\small\textcolor{blue}{(69.8\%)}} \\
\hline
\rowcolor{gray!15}
\multicolumn{4}{c}{\textbf{GPT-4o}} \\
\hline
\textbf{MARA} & \textbf{84.64} {\small\textcolor{blue}{(96.1\%)}} & \textbf{68.02} {\small\textcolor{blue}{(93.6\%)}} & \textbf{32.98} {\small\textcolor{blue}{(74.5\%)}} \\
w/o SEC & 81.22 {\small\textcolor{blue}{(92.2\%)}} & 62.35 {\small\textcolor{blue}{(85.8\%)}} & 32.18 {\small\textcolor{blue}{(72.7\%)}} \\
w/o QRE & 79.74 {\small\textcolor{blue}{(90.5\%)}} & 59.43 {\small\textcolor{blue}{(81.8\%)}} & 31.85 {\small\textcolor{blue}{(72.0\%)}} \\
w/o both & 78.21 {\small\textcolor{blue}{(88.8\%)}} & 58.27 {\small\textcolor{blue}{(80.2\%)}} & 30.12 {\small\textcolor{blue}{(68.1\%)}} \\
\hline
\end{tabular}}
\end{table}

\noindent\textbf{Overall Framework Ablation.}
We first examine the contribution of each module in the MARA framework by progressively removing the QRE and the SEC. As shown in Table~\ref{table3}, \textit{w/o SEC} denotes replacing SEC with a static top-$3$ strategy, while \textit{w/o QRE} removes the retrieval-side granularity-aware encoder. \textit{w/o Both} reduces the pipeline to a standard visual-language RAG baseline.

Removing either module leads to notable drops in performance, confirming that both QRE and SEC independently contribute to generation quality. The largest degradation occurs when both are removed, highlighting the synergy between retrieval and generation adaptivity. Interestingly, SEC yields greater gains under GPT-4o compared to MiniCPM, suggesting that stronger generators benefit more from dynamic evidence control.

\begin{table}[htbp]
\centering
\caption{Ablation study of QRE. The best performance in each column is highlighted in \textbf{bold}.}
\label{table4}
\resizebox{0.75\linewidth}{!}{
\begin{tabular}{l|c|c|c}
\hline
\textbf{Method} & \textbf{DocVQA} & \textbf{InfoVQA} & \textbf{PlotQA} \\
\hline
\rowcolor{gray!15}
\multicolumn{4}{c}{\textbf{(a) HAR (HF)}} \\
\hline
\textbf{QRE} & \textbf{81.69} & \textbf{86.99} & 39.90 \\
w/o fine & 80.36 & 86.01 & 40.15 \\
w/o coarse & 80.98 & 86.45 & \textbf{40.21} \\
w/o both & 77.64 & 84.12 & 40.19 \\
\hline
\rowcolor{gray!15}
\multicolumn{4}{c}{\textbf{(b) HAR (Our Rep.)}} \\
\hline
\textbf{QRE} & \textbf{83.35} & \textbf{87.93} & 43.23 \\
w/o fine & 82.25 & 86.80 & 43.37 \\
w/o coarse & 82.63 & 87.14 & \textbf{43.58} \\
w/o both & 79.82 & 84.61 & 43.55 \\
\hline
\end{tabular}
}
\end{table}

\noindent\textbf{Granularity and Fusion in QRE.}
We further analyze QRE by ablating its internal design components. As shown in Table~\ref{table4}, we compare variants using only coarse-grained (\textit{w/ Coarse}) or fine-grained (\textit{w/ Fine}) features, and a naive combination without query-adaptive fusion. Neither single-level encoder achieves competitive results, and the naive fusion fails to balance relevance and context. In contrast, our full QRE leverages hierarchical document structure: coarse features provide layout-level global priors, while fine features recover localized details. The query-aware fusion mechanism effectively reweights their contributions, yielding more precise retrieval across heterogeneous document types.

\subsection{Analysis and Discussion}

\subsubsection{Balancing Input Efficiency and Answer Quality}
\begin{figure}
    \centering
    \includegraphics[width=0.9\linewidth]{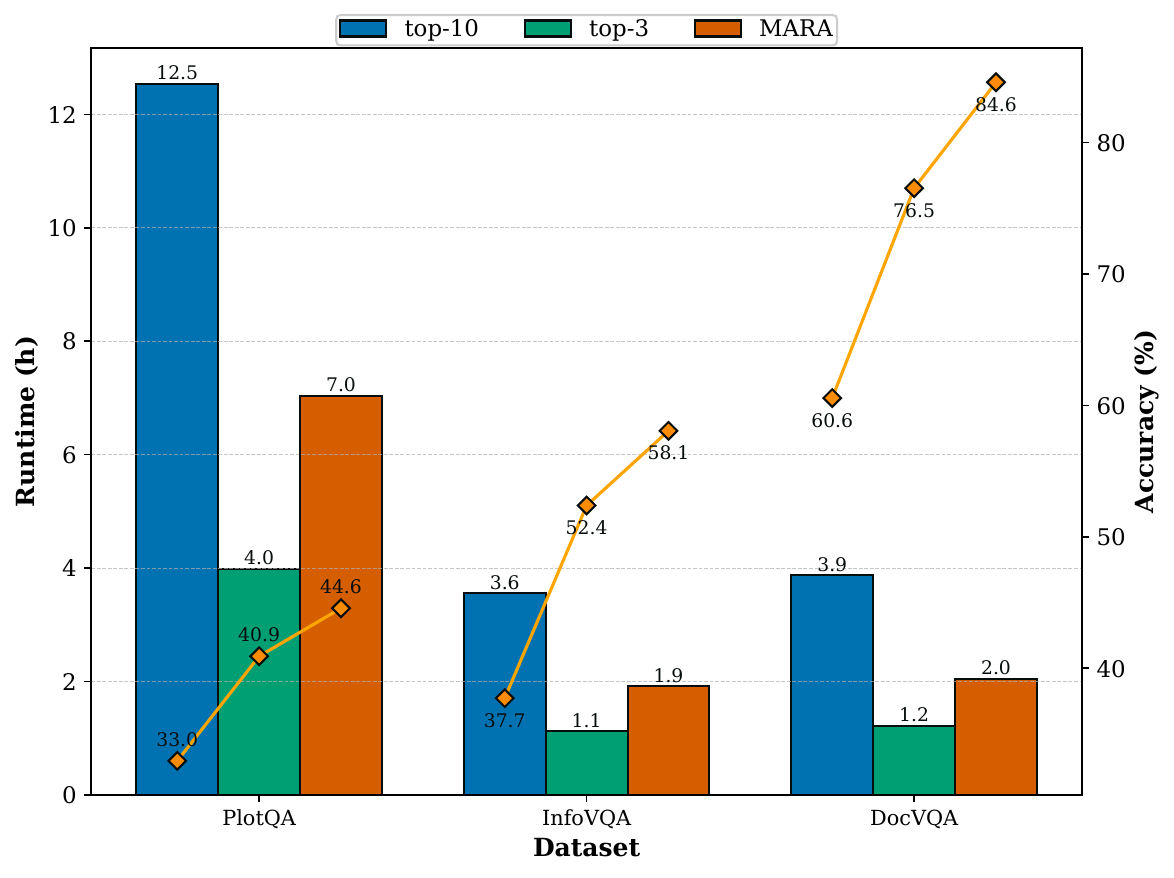}
    \caption{Comparison of runtime and accuracy across input strategies. MARA achieves higher accuracy than static top-$k$ baselines with significantly reduced runtime, demonstrating a superior efficiency-accuracy tradeoff.}
    \label{fig3}
\end{figure}

Figure~\ref{fig3} compares the runtime and generation accuracy of static top-$k$ strategies (top-3, top-10) and our proposed method MARA across three representative datasets. While top-10 achieves moderate accuracy, it suffers from substantial runtime overhead due to processing a large number of documents, especially on complex tasks such as PlotQA. In contrast, top-3 is more efficient but often misses critical evidence, leading to lower answer quality.

MARA achieves the highest accuracy across all datasets while significantly reducing runtime compared to top-10 retrieval. This efficiency gain stems from the Self-Reflective Evidence Controller (SEC), which expands the input context incrementally and only when necessary. By dynamically assessing evidence sufficiency, SEC prevents generator overload while retaining informative content. As a result, MARA achieves a favorable efficiency–accuracy tradeoff, demonstrating the advantage of adaptive evidence integration over static input expansion.

\subsubsection{SEC Iteration Behavior}
To evaluate the runtime behavior of the Self-Reflective Evidence Controller (SEC), we report call statistics across datasets in Table~\ref{table5}. On average, the controller requires fewer than 2 calls per query (1.90 overall). PlotQA shows the highest average (2.31), reflecting the need for iterative reasoning in complex visual-numerical scenarios, whereas simpler datasets such as SlideVQA (1.15) and InfoVQA (1.12) converge with minimal evidence expansion.

These results confirm that SEC effectively regulates context usage without incurring excessive generation cost. Even on challenging datasets, the average iteration count remains below static top-$k$ strategies (e.g., $k=10$), validating its efficiency–accuracy tradeoff.

\begin{table}[htbp]
\centering
\caption{Call statistics across different datasets.}
\label{table5}
\resizebox{0.85\linewidth}{!}{
\begin{tabular}{c|ccc}
\hline
\textbf{Dataset} & \textbf{PlotQA} & \textbf{ArXivQA} & \textbf{SlideVQA} \\
\hline
\textbf{Total Calls} & 26082 & 13020 & 1889 \\
\textbf{Queries}     & 11307 & 8640  & 1640 \\
\textbf{Avg. Calls}  & 2.31  & 1.51  & 1.15 \\
\hline
\textbf{Dataset} & \textbf{InfoVQA} & \textbf{ChartQA} & \textbf{DocVQA} \\
\hline
\textbf{Total Calls} & 2287 & 1472 & 2519 \\
\textbf{Queries}     & 2046 & 718  & 1879 \\
\textbf{Avg. Calls}  & 1.12 & 2.05 & 1.34 \\
\hline
\textbf{Overall Avg. Calls} & \multicolumn{3}{c}{\textbf{1.83}} \\
\hline
\end{tabular}
}
\end{table}

\subsubsection{Impact of Attention Temperature}
\begin{figure}
    \centering
    \includegraphics[width=0.9\linewidth]{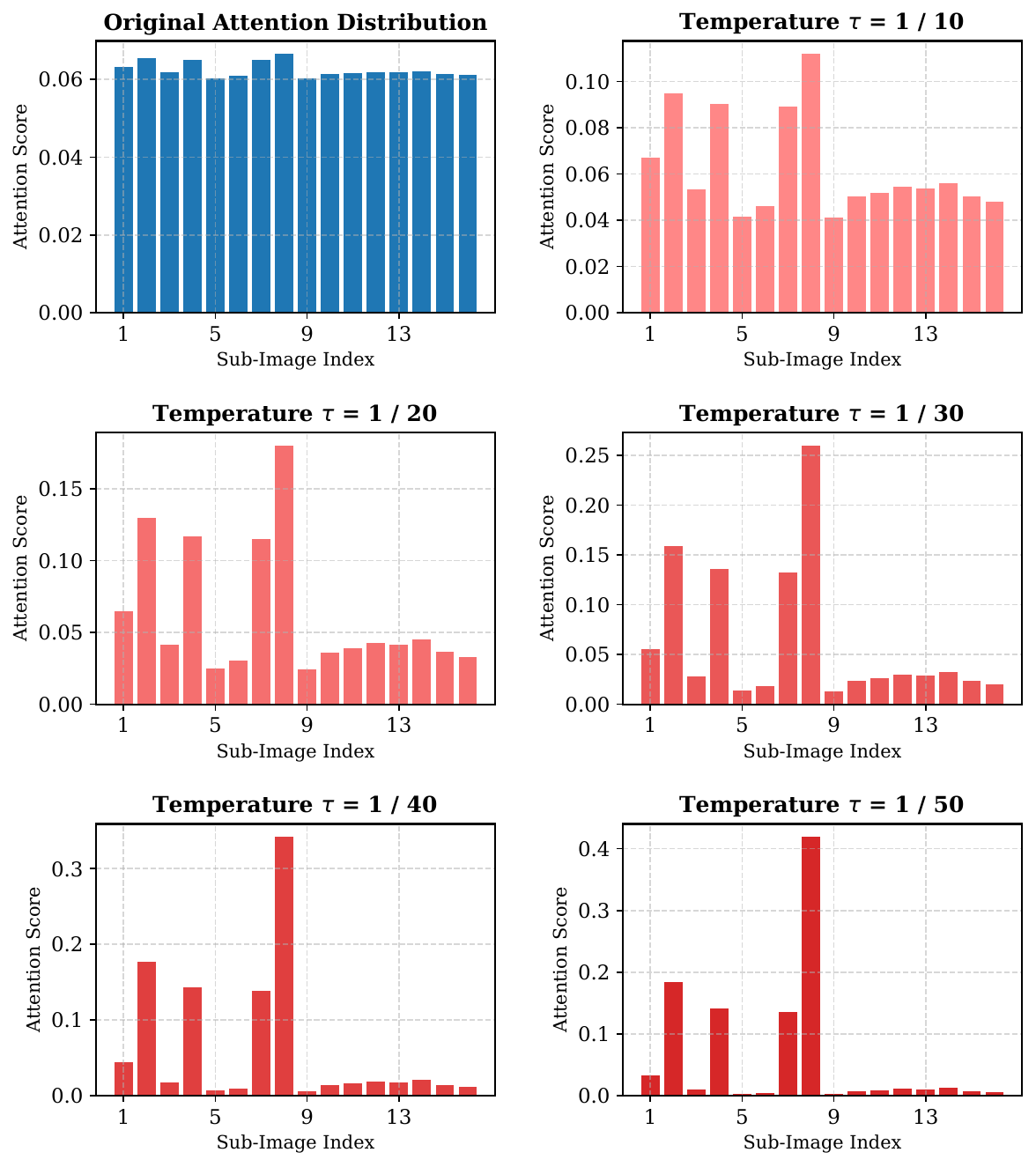}
    \caption{Query-to-region attention distributions under different temperature values $\tau$. A moderate value (e.g., $\tau=1/20$) produces more focused yet balanced attention.}
    \label{fig4}
\end{figure}

We analyze the impact of the temperature parameter $\tau$ in the query-aligned attention mechanism, which adjusts the sharpness of attention over document regions. As shown in Figure~\ref{fig4}, a high $\tau$ (e.g., $1/10$) produces nearly uniform scores, reducing the model’s ability to distinguish relevant regions, while a very low $\tau$ (e.g., $1/50$) creates overly concentrated attention that may overlook weak but important signals. Empirically, a moderate setting (e.g., $1/20$) offers the best trade-off, enabling focus on key regions while preserving context. This validates our design of a learnable $\tau$ that adapts attention concentration to query-document interactions.

Further analyses, including the impact of subregion granularity and additional ablation studies, are provided in the Appendix.

\section{Conclusion}

We propose MARA, a novel framework for retrieval-augmented generation on multimodal documents, addressing the core limitations of query-agnostic encoding and static evidence selection. MARA introduces two key components: the Query-Aligned Region Encoder (QRE), which performs hierarchical, query-aware retrieval across global, coarse, and fine-grained document structures; and the Self-Reflective Evidence Controller (SEC), which enables the generator to iteratively assess and expand input context based on sufficiency feedback. Extensive experiments on six diverse document QA benchmarks demonstrate that MARA consistently outperforms strong multimodal baselines in both retrieval and generation.Detailed analysis further reveals that our adaptive retrieval and generation strategies yield higher accuracy with controllable computational cost. We also visualize attention distributions and controller behaviors to highlight the interpretability and effectiveness of our design. Overall, MARA offers a unified and flexible approach to dynamic multimodal reasoning, and provides new insights into document-level retrieval-generation synergy.

\section*{Acknowledgments}
The work is supported by the Key Laboratory of Target Cognition and Application Technology under Grant (2023-CXPT-LC-005).

\bibliography{sample-base}

\clearpage
\onecolumn
\appendix

\section{Experiment Details}

\subsection{Datasets }\label{appendix-datasets}
\begin{table}[H]
\centering
\caption{Statistics of document QA datasets.}
\begin{tabular}{lccc}
\toprule
\textbf{Dataset} & \textbf{\#Queries} & \textbf{\#Images} & \textbf{\#Pos. Docs/Query} \\
\midrule
SlideVQA & 1,640 & 1,284 & 1.34 \\
ArXivQA & 8,640 & 8,066 & 1.00 \\
ChartQA & 718 & 500 & 1.00 \\
MP-DocVQA & 1,879 & 741 & 1.00 \\
InfoVQA & 2,046 & 459 & 1.00 \\
PlotQA & 11,307 & 9,593 & 1.00 \\
\bottomrule
\end{tabular}
\label{table6}
\end{table}

\subsection{Hyper Parameters}\label{appendix-hyper}
\begin{table}[H]
    \centering
    \label{table7}
    \caption{Granularity fusion weights and attention scaling factors across datasets. All values are set from a small search space, and most tasks share similar configurations, indicating stable cross-domain behavior.}
    \resizebox{0.7\linewidth}{!}{
    \begin{tabular}{c|c|c|c|c|c|c}
    \hline
    \rowcolor{gray!15}
    \textbf{H. Param.}
    & \textbf{ArxivQA}
    & \textbf{ChartQA}
    & \textbf{DocVQA}
    & \textbf{InfoVQA}
    & \textbf{PlotQA}
    & \textbf{SlideVQA} \\
    \hline
    $g_c$       & 0.2  & 0.1  & 0.2  & 0.2  & 0.1  & 0.2 \\
    $1/\tau_c$   & 20   & 15   & 30   & 40   & 10   & 30  \\
    $g_f$       & 0.1  & 0.2  & 0.2  & 0.2  & 0.1  & 0.2 \\
    $1/\tau_f$   & 20   & 20   & 20   & 50   & 10   & 30  \\
    \hline
    \end{tabular}
    }
\end{table}

\section{Recall}
\begin{table*}[htbp]
\centering
\caption{Ablation study of HAR on Recall@10. Improvements are shown with arrows.}
\label{table8}
\resizebox{0.85\linewidth}{!}{
\begin{tabular}{l|l|l|l|l|l|l}
\hline
\textbf{Model / Method} & \textbf{ArXivQA} & \textbf{ChartQA} & \textbf{DocVQA} & \textbf{InfoVQA} & \textbf{PlotQA} & \textbf{SlideVQA} \\
\hline
\rowcolor{gray!15}
\multicolumn{7}{c}{\textbf{(a) HAR (HF)}} \\
\hline
Base & 80.42 & 72.42 & 92.92 & 96.24 & 61.40 & 96.86 \\
w/ coarse & 81.94 (↑1.52) & 73.40 (↑0.98) & 94.41 (↑1.49) & 97.46 (↑1.22) & 61.74 (↑0.34) & 97.45 (↑0.59) \\
w/ fine & 81.48 (↑1.06) & 74.09 (↑1.67) & 95.58 (↑2.66) & 97.51 (↑1.27) & 61.37 (↓0.03) & 97.51 (↑0.65) \\
w/ coarse \& fine & 81.99 (↑1.57) & 73.96 (↑1.54) & 95.16 (↑2.24) & 97.61 (↑1.37) & 61.16 (↓0.24) & 97.60 (↑0.74) \\
\hline
\rowcolor{gray!15}
\multicolumn{7}{c}{\textbf{(b) HAR (Our Rep.)}} \\
\hline
Base & 83.81 & 72.98 & 94.09 & 96.48 & 64.03 & 97.06 \\
w/ coarse & 84.97 (↑1.16) & 74.51 (↑1.53) & 95.64 (↑1.55) & 97.75 (↑1.27) & 63.97 (↓0.06) & 97.43 (↑0.37) \\
w/ fine & 84.86 (↑1.05) & 75.21 (↑2.23) & 96.06 (↑1.97) & 97.90 (↑1.42) & 63.97 (↓0.06) & 97.64 (↑0.58) \\
w/ coarse \& fine & \textbf{85.22} (↑1.41) & \textbf{75.63} (↑2.65) & \textbf{96.59} (↑2.50) & \textbf{98.04} (↑1.56) & 63.87 (↓0.16) & \textbf{97.73} (↑0.67) \\
\hline
\end{tabular}
}
\end{table*}

\newpage
\section{Prompts}
\label{sec:appendix}

\begin{mdframed}[
    frametitle={Prompt for Evaluating Information Sufficiency},
    frametitlerule=true,
    frametitlefont=\color{white}\bfseries,
    frametitlebackgroundcolor=gray!70,
    linewidth=1.2pt,
    roundcorner=4pt, 
    backgroundcolor=gray!10, 
    linecolor=black, 
    align=center, 
    userdefinedwidth=0.85\textwidth,
]
You are a fact-based reasoning assistant. Your goal is to assess whether the provided information is sufficient to generate an accurate and complete answer to the given question.

\#\#\# User Question:
\{user\_question\}

\#\#\# Context Provided (Retrieved Passages + Memory Buffer):
\{retrieved\_passages\}

\#\#\# Instructions:

- If the provided information is **fully sufficient** to generate a factually accurate and complete answer, return: "Sufficient"

- If the information **contains some relevant content but is missing critical details**, return: "Partially Sufficient"

- If the information is **missing key facts or is entirely insufficient**, return: "Insufficient"

\#\#\# Response Format (Only output one word: "Sufficient", "Partially Sufficient", or "Insufficient"):
\end{mdframed}

\vspace{4ex}

\begin{mdframed}[
    frametitle={Feedback Generation Prompt},
    frametitlerule=true,
    frametitlefont=\color{white}\bfseries,
    frametitlebackgroundcolor=gray!70,
    linewidth=1.2pt,
    roundcorner=4pt, 
    backgroundcolor=gray!10, 
    linecolor=black, 
    align=center, 
    userdefinedwidth=0.85\textwidth,
]
You are a retrieval-augmented reasoning expert. Your task is to analyze whether the provided context contains enough information to generate a complete and accurate answer. If not, generate structured feedback to guide the next retrieval step.

\#\#\# User Question:
\{user\_question\}

\#\#\# Context Provided (Retrieved Passages + Memory Buffer):
\{retrieved\_passages\}

\#\#\# Sufficiency Assessment:
\{previous\_sufficiency\_result\}  \# "Partially Sufficient" or "Insufficient"

\#\#\# Instructions:

- Identify **missing key details** needed to fully answer the question.

- Suggest **which aspects should be retrieved in the next step** (e.g., missing facts, numerical data, explanations).

- If the context contains **irrelevant information that can be ignored**, list what should be **excluded** in the next retrieval step.

- Your response should be **concise and structured** in the following format:

\#\#\# Response Format:

- **Missing Information:** [List missing aspects or facts]

- **Retrieval Guidance:** [Suggest which type of content to retrieve next]

- **Irrelevant Content:** [List any unnecessary or misleading information]
\end{mdframed}

\newpage
\begin{mdframed}[
    frametitle={Memory Update Prompt},
    frametitlerule=true,
    frametitlefont=\color{white}\bfseries,
    frametitlebackgroundcolor=gray!70,
    linewidth=1.2pt,
    roundcorner=4pt, 
    backgroundcolor=gray!10, 
    linecolor=black, 
    align=center, 
    userdefinedwidth=0.85\textwidth,
]
You are managing a retrieval memory buffer for an adaptive reasoning system. Your task is to determine which retrieved passages should be stored in long-term memory and which should be discarded.

\#\#\# User Question:
\{user\_question\}

\#\#\# Current Memory Buffer:
\{current\_memory\}

\#\#\# Newly Retrieved Passages:
\{retrieved\_passages\}

\#\#\# Instructions:

- **Keep** passages that contain **unique, critical, and factual information** necessary to answer the question.

- **Remove** passages that are **redundant, irrelevant, or contradict existing memory**.

- **Prioritize** storing passages that provide **clarifications, numerical data, or supporting evidence**.

- **If memory is full**, remove **the least relevant passage** to make space.

\#\#\# Response Format:

- **Keep in Memory:** [List passages to retain]

- **Remove from Memory:** [List passages to discard]
\end{mdframed}

\end{document}